# Computing trends using graphic processor in high energy physics

Mihai Niculescu[1,2], Sorin-Ion Zgura[2]

1 – Faculty of Physics, University of Bucharest,

Bucharest, Romania

mihai@spacescience.ro

2 - Institute of Space Sciences, ISS

Magurele, Ilfov, Romania

szgura@spacescience.ro

*Abstract*—One of the main challenges in Heavy Energy Physics is to make fast analysis of high amount of experimental and simulated data. At LHC-CERN one p-p event is approximate 1 Mb in size. The time taken to analyze the data and obtain fast results depends on high computational power. The main advantage of using GPU(Graphic Processor Unit) programming over traditional CPU one is that graphical cards bring a lot of computing power at a very low price. Today a huge number of application(scientific, financial etc) began to be ported or developed for GPU, including Monte Carlo tools or data analysis tools for High Energy Physics. In this paper, we'll present current status and trends in HEP using GPU.

## I. Introduction

### A. Computing status in High Energy Physics

Today experiments from High Energy Physics (HEP) all around the world are dealing with challenging new horizons. As an example, the Large Hadron Collider (LHC) from European Center for Nuclear Research (CERN) does not only drive forward theoretical, experimental and detector physics but also pushes to limits computing. In this sense, it is estimated that every month 2 Petabytes of data must be stored and analyzed off-line.

The complexity of the computing challenges and required computing resources made HEP experiments to create their own computing model.

In this computing model, partner institutions combine computer resources in a distributed system in order to store and process data. This is known as GRID computing. The structure and software used is unique for every experiment. Beside grid computing a lot of effort is involved in the development of simulations and analysis codes: Monte Carlo simulation tools data processing frameworks, analysis algorithms, detector simulations, etc.

The performance of the simulations or analysis is determined by the methods and optimizations involved in development of software. But ultimately, the time until searched results are obtained is determined by computing resources, more specifically: hardware.

*B. Possible Solution: Graphic Cards*

Using graphic cards for scientific computing is not a new idea and it has been around before graphic processor (GPU) creators decided to publish Application Programming Interfaces (API). Driven by the gaming industry, graphic cards and software libraries have been developed to use GPUs as a general-purpose computing device (GPGPU). The primary GPGPU providers are NVIDIA and AMD/ATI. Most graphics cards from NVIDIA can be programmed using Compute Unified Device Architecture (CUDA) an API extension to C language.

Currently, AMD/ATI focus on developing OpenCL (Open Computing Language), a GPGPU software developed in consortium at Khronos Group. OpenCL goal is to be a cross platform GPGPU toolkit; this means, using same API, the GPU application can be compiled and run on processors from NVIDIA, ATI, CELL or S3 graphic cards.

In this paper, we present the API from NVIDIA for GPUs: CUDA and its current utilization in High Energy Physics.

## II. CUDA

CUDA [1] - Compute Unified Device Architecture- is a technology developed by NVIDIA that allows programmers to use a language based on C (C with extensions) to write applications for execution on GPU. Although, CUDA comes with an API for C language, bindings for other languages are supported (C++, FORTRAN, RUBY, Python, etc).

Freely available from NVIDIA site, CUDA comes with a Toolkit and SDK (Software Development Toolkit). The CUDA Toolkit contains software and documentation necessary to build GPU applications: nvcc C compiler, libraries for fast Fourier transformation (CUFFT) and basic linear algebra (CUBLAS), a cuda profiler, etc. Optionally, one can install the CUDA SDK which contains lots of examples and makes easy learning and development.

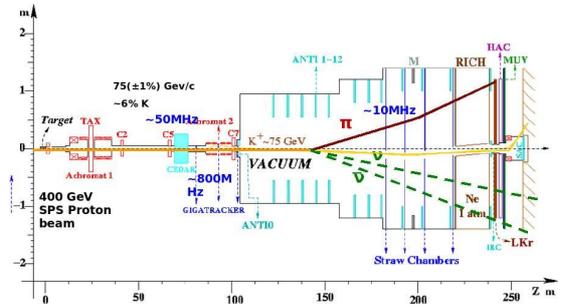

Figure 1. NA62 Experiment

The execution model is SIMT (Single Instruction Multiple Threads). One function (named kernel) is executed N times in parallel by N different CUDA threads. These threads are automatically managed by CUDA and such does not require explicit management. However, developers must analyze data structure and determine how to divide data into small chunks for distribution among the threads. Such, one of the main job of a CUDA developer is to analyze memory access and distribution, due to the fact GPU memory is limited and much smaller than CPU memory (RAM).

## III. CUDA AT CERN

*A. NA62 Experiment [2]*

The NA62 experiment (Fig. 1) from CERN (European Organization for Nuclear Reasearch, near Geneva) aims at measuring the Branching Ratio of $K+ \rightarrow \pi + \nu\ \nu$, predicted in the standard model at level $10^{-10}$. The trigger system is a very crucial part for the experiment. The reduction of interesting data has to be very effective in order to decrease the total bandwidth requirements for the data readout.

The usage of graphic card processing in the construction of a trigger system, both in hardware and software, allowed to make decisions in real-time and increased the computing capability in the trigger.

To test the performances and feasibility, three algorithms were implemented. The tests were done on NVIDIA Tesla hardware with one GPU GT200 containing 240 cores, 4 GB DDR3 memory with 800 MHz and a bandwidth of 102 GB/s.

The NA62 experiment concluded that the ratio of cost over performance is very appealing and the new

TABLE I. PERFORMANCES OF NA62 ALGORITHMS ON GPU

| Algorithm | Time (μs) |
|-----------|-----------|
| GHT | 85 |
| OPMH | 139 |
| ODMH | 12.4 |

TABLE II. TMATRIX WITH CUDA. EXECUTION TIME OF MATRIX MULTIPLICATION FOR DIFFERENT MATRIX DIMENSIONS

| Matrix Dimension | Time Taken in CPU (in seconds) | Time taken in GPU (in seconds) |
|------------------|-------------------------------|-------------------------------|
| 1024X1024 | 4.0 | 0.52 |
| 2048X2048 | 130.62 | 1.51 |
| 3072X3072 | 490.98 | 5.25 |
| 4096X4096 | 1270.22 | 9.94 |

hardware and software scheme implied by graphic cards can be adapted for high energy physics.

## A. Integration of CUDA in AliRoots [3]

The AliRoot framework is an object-oriented simulation, reconstruction and data analysis framework based on ROOT and Virtual Monte Carlo interface. It includes core services for detector simulations and off line processing and analysis of the from the detectors.

The integration of CUDA in AliRoot is done in 2 steps:

- **Building System:** Using FindCuda.cmake – CUDA is integrated in AliRoot building system very smoothly.
- **AliCuda-ROOT interface:** An interface is implemented which permits the use of GPU's implemented function from within a ROOT CINT session.

Tests on matrix multiplication were performed on a Nvidia GeForce 8400 GS with a host processor Intel Core 2 Duo CPU E8400 (3.0 GHz) with 2GB DDR2 SDRAM. Results of tests are found in Table 2.

## IV. CUDA AT FAIR/GSI

### A. Integration of CUDA in FairRoot [4]

FairRoot is the simulation and data analysis framework used at FAIR (Facility for Antiproton and Ion Research, Darmstadt, Germany).

CUDA is integrated in FairRoot same as in AliRoot, in two steps:

- **A Building System** Using FindCuda.cmake The users do not have to take care of Makefiles or which compiler should be called (e.g. NVCC or GCC).
- **FairCuda:** An interface is implemented which enables the use of GPU's implemented function from within a ROOT CINT session. The CUDA implemented kernels are wraped by a class (FairCuda) that is implemented in ROOT and has a dictionary. From a ROOT CINT session the user simply call the wraper functions which call the GPU functions (kernels).

### B. PANDA Experiment [5]

One of track fitting algorithms used by PANDA experiment (GSI Darmstadt) was ported to CUDA. The algorithm ported to CUDA uses tasks from FairCuda interface.

Different tests were performed on CPU and GPU in different modes: single float precision, double precision, and emulation mode - special mode in CUDA to emulate application running on CPU, see Table 3.

Using GPUs for track fitting one can win orders of magnitudes in performance compared to the CPUs, however one has to determine how to divide the data into smaller chunks for distribution among the thread processors (GPUs).

TABLE III. TRACK FITTING AT PANDA EXPERIMENT. A GAIN OF ALMOST 70 TIMES ON GPU VS CPU

| Mode/Track/Event | 50 | 100 | 1000 | 2000 |
|------------------|-----|------|------|------|
| GPU (Emu) | 6.0 | 15.0 | 180 | 370 |
| CPU | 3.0 | 5.0 | 120 | 230 |
| GPU (Double) | 1.2 | 1.5 | 3.2 | 5.0 |
| GPU (Float) | 1.0 | 1.2 | 1.8 | 3.2 |

## C. CBM Experiment [6]

The CBM (Condensed Barionic Matter) Experiment at FAIR is a dedicated fixed target heavy ion experiment.

The Kalman filter algorithm from CBM experiment was parallelized for Intel SSE and CELL using SIMD (Single Instruction Multiple Data) model. A speed up of 10000 times was determined for the SIMD version against the initial implementation. Porting the algorithm to GPU was the logical next step. [7]

Porting the modified algorithm to GPU required noticing the differences between the two architectures: SIMD and SIMT, which is found on GPU hardware. GT200 chip contains 30 of these multiprocessors.

In contrast to the single instruction multiple data (SIMD) based architectures of the NVIDIA GPUs are based on a single instruction multiple threads (SIMT) model, which means each ALU is connected with an own instruction counter. Therefore the algorithm needs to be parallelized on a thread level instead of the below as in the SIMD case.

Comparison results are found in Table 4.

TABLE IV. CBM KALMAN FILTER PERFORMANCE ON DIFFERENT DEVICES. HIGH VALUE OF COLUMN TRACKS/TIME MEAN BETTER PERFORMANCE.

| Device | Clock speed (GHz) | Tracks/time ($10^6$/s) |
|---|---|---|
| Intel Xeon (1 core) | 2.66 | 0.680 |
| AMD Opteron (1 core) | 1.8 | 0.538 |
| Cell SPE (1SPE) | 2.4 | 1.15 |
| Intel Core 2 (1 core) | 2.4 | 1.47 |
| NVIDIA 8800 GTS 512 | 1.6 | 13.0 |
| NVIDIA GTX 280 | 1.3 | 21.7 |

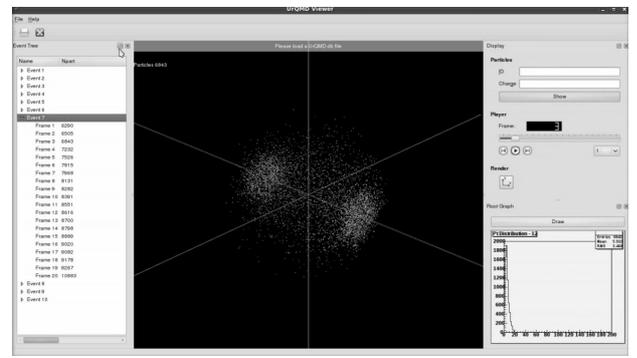

Figure 2. HEPLab Application main window. File Structure is represented in left panel(Events, time frames) in the simulation. In the center, one can see the 3D nuclear interaction. In the right side, there is a particle filter at top, a player controls for the animation. At bottom right side, distribution of the total transversal momentum.

## V. CUDA AT ISS [8]

CUDA has been an interested subject for the Computing group, in ISS (Institute of Space Sciences, Magurele, Romania) almost since its first public release.

We searched for new possibilities of using GPU technology in scientific computing. As such, we developed an application with usage in high energy physics and education: HEPLab Viewer.

HEPLab application contains a GUI (Graphical User Interface) which allows the 3D visualization of the nuclear interaction, while in the same time computing the transversal momentum, pseudo-rapidity and generating distributions of these, see Fig. 2.

The hardware used was NVIDIA graphic board GeForce 9600 GT (8 processors) with 1GB DRAM at 900MHz and 64 cores at 1.6GHz. A comparative test was done on cluster of 16 CPUs CORE 2 DUO at 2.2GHz and 2GB RAM per core.

The benchmark results are in Table 5.

## VI. CONCLUSION

| interaction | GeForce 9600 GT (MB/seconds) | 16 Intel Core2Duo 2.2GHz (MB/seconds) |
|---|---|---|
| proton-proton | 30 | 4 |
| Gold-Gold | 37 | 6 |

CUDA technology permits writing GPU application using familiar programming concepts.

Orders of magnitude can be obtained if one succeeds much parallelization of its algorithm and optimization of memory distribution and access.

Few HEP algorithms have been ported to GPU, but some may not even be possible to parallelize. Understanding different device memory regions (GPU RAM) is crucial to getting better performance and simplify problems.

Future graphic cards (GPUs) will bring more computing horse-power with performance improvements and new challenges to find the optimum balance between parallelization and efficient memory usage.

ACKNOWLEDGMENT

The participation and work of MN at The 18-th FCAL Collaboration was supported by European Space Agency through PECS RoSpaceGRID C98050 and University of Bucharest, Faculty of Physics through program Burse Doctorale from Romanian Ministry of Labor, Family and Social Protection POSDRU/88/1.5/S/56668.

REFERENCES

[1] NVIDIA http://developer.nvidia.com/object/cuda.html
[2] Gianluca Lamanna, Gianmaria Collazuol and Marco Sozzi , GPUs for fast triggering and pattern matching at the CERN experiment NA62 , http://www.sciencedirect.com/science/article/pii/S0168900210022011
[3] Subhankar Biswas , GPUs for accelerating compute-intensive tasks in the AliRoot Framework , CERN Summer School 2010, http://indico.cern.ch/getFile.py/access?contribId=3&resId=0&materialId=slides&confId=99464
[4] M. Al-Turany , F. Uhlig , R. Karabowicz , GPU's for event reconstruction in the FairRoot
[5] Framework , Journal of Physics: Conference Series 219 (2010) 042001 , doi:10.1088/1742-6596/219/4/042001
[6] M. Bach, S. Gorbunov, I. Kisel , V. Lindenstruth, and U. Kebschull, Porting a Kalman filter based track fit to NVIDIA CUDA , GSI SCIENTIFIC REPORT 2008 , http://www.gsi.de/scirep2008/PAPERS/FAIR-EXPERIMENTS-38.pdf
[7] Matthias Bach , SIMT Kalman Filter - High throughput track fitting , 2010-06-11 , https://www.gsi.de/documents/DOC-2010-Jun-125-1.pdf
[8] A.JIPA , C.M. MITU, M. NICULESCU,S.I ZGURA, DATA ANALYSIS AND 3D EVOLUTION IN HIGH ENERGY PHYSICS USING GRAPHIC PROCESSOR , Conferinta Nationala de Fizica 2010, Iasi,